\documentclass[11pt,a4paper]{article}
\usepackage{amssymb}
\usepackage{amsmath}\usepackage{graphicx}  
\numberwithin{equation}{section}
\usepackage{footnote}
\makesavenoteenv{tabular}

\newcommand{\id}[1]{\ensuremath{\mathrm{id}}}

\newcommand{\hi}[1]{\emph{\textbf{#1}}}
\newcommand{\qm}{quantum mechanics}
\newcommand{\er}{\eqref}
 
\newcommand{\beq}{\begin{equation}}
\newcommand{\eeq}{\end{equation}} 
\newcommand{\bea}{\begin{eqnarray}}
\newcommand{\eea}{\end{eqnarray}} \newcommand{\nn}{\nonumber}

\newcommand{\ul}{\underline}

 \newcommand{\til}{\tilde}
\newcommand{\raw}{\rightarrow}

 \newcommand{\Raw}{\Rightarrow}

\newcommand{\ot}{\otimes} 
\newcommand{\la}{\langle} \newcommand{\ra}{\rangle}

\newcommand{\x}{\times}

\newcommand{\ca}{C*-algebra}

\newcommand{\Hs}{Hilbert space}

\newcommand{\hv}{hidden variable}

\newcommand{\al}{\alpha} 
 
\newcommand{\dl}{\delta} \newcommand{\Dl}{\Delta}

\newcommand{\lm}{\lambda} \newcommand{\Lm}{\Lambda}
\newcommand{\rh}{\rho} \newcommand{\sg}{\sigma}
\newcommand{\Sg}{\Sigma} \newcommand{\ta}{\tau} 
 \newcommand{\phv}{\varphi}
\newcommand{\ch}{\ch} \newcommand{\ps}{\psi} 
\newcommand{\om}{\omega} \newcommand{\Om}{\Omega}

\newcommand{\Tr}{\mbox{\rm Tr}\,}

 \newcommand{\CP}{{\mathcal P}}

\newcommand{\C}{{\mathbb C}} 
\newcommand{\N}{{\mathbb N}} \newcommand{\R}{{\mathbb R}}
 
  \makeatletter
 
\makeatletter
\def\moverlay{\mathpalette\mov@rlay}
\def\mov@rlay#1#2{\leavevmode\vtop{%
   \baselineskip\z@skip \lineskiplimit-\maxdimen
   \ialign{\hfil$\m@th#1##$\hfil\cr#2\crcr}}}
\newcommand{\charfusion}[3][\mathord]{
    #1{\ifx#1\mathop\vphantom{#2}\fi
        \mathpalette\mov@rlay{#2\cr#3}
      }
    \ifx#1\mathop\expandafter\displaylimits\fi}
\makeatother

\newtheorem{definition}{definition}[section]

\newtheorem{theorem}[definition]{Theorem}
\newtheorem{proposition}[definition]{Proposition}

\newtheorem{corollary}[definition]{Corollary}
\topmargin = - 1 cm \textheight = 23 cm \textwidth = 15 cm
\begin{document} \renewcommand{\thefootnote}{\alph{footnote}}
\pagenumbering{arabic} \setlength{\unitlength}{1cm}\cleardoublepage
\date\nodate
\begin{center}
\begin{huge}
{\bf Indeterminism and Undecidability}\end{huge}
\bigskip\bigskip

\begin{Large}
 Klaas Landsman\vspace{5mm}
 \end{Large}
 
 \begin{large}
 Department of Mathematics, 
Institute for Mathematics, Astrophysics, and \\ Particle Physics (IMAPP), Radboud University, Nijmegen, The Netherlands
Email:
\texttt{landsman@math.ru.nl}
\end{large}
\smallskip
\begin{center}
\emph{Dedicated to the memory of Michael Redhead (1929--2020)}
\end{center}
 \begin{abstract} 
\noindent 
The aim of this paper is to argue that the (alleged) indeterminism of \qm, claimed by adherents of the Copenhagen interpretation  since Born (1926), can be proved from Chaitin's follow-up to G\"{o}del's (first) incompleteness theorem. 
In comparison,  Bell's (1964) theorem as well as the so-called free will theorem--originally due to 
 Heywood and Redhead (1983)--left two loopholes for deterministic  \hv\ theories, namely giving up either \emph{locality} (more precisely:  local contextuality, as in Bohmian mechanics) or  \emph{free choice} (i.e.\ uncorrelated measurement settings, as in 't Hooft's cellular automaton interpretation of \qm).  
 The main point is that Bell and others did not exploit the full empirical content of quantum mechanics, which consists of  long series of outcomes of repeated measurements (idealized as infinite binary sequences): their arguments only used the long-run relative frequencies derived from such series, and hence merely asked \hv\ theories to reproduce  single-case Born probabilities defined by certain entangled bipartite states. If we idealize binary outcome strings  of a  fair quantum coin flip as infinite sequences, \qm\ predicts that these  typically (i.e.\ almost surely) have a  property called  \emph{1-randomness} in logic, which is much stronger than 
uncomputability. This is the key to my claim, which is admittedly based on a stronger (yet compelling) notion of determinism than what is common in the  literature on \hv\ theories.
\end{abstract}\end{center}
\tableofcontents
\bigskip

\noindent\makebox[\linewidth]{\rule{\textwidth}{0.4pt}}

\thispagestyle{empty}
\renewcommand{\thefootnote}{\arabic{footnote}}
\newpage \setcounter{footnote}{0}
\section{Introduction: G\"{o}del and Bell}
 While \emph{prima facie} totally unrelated, G\"{o}del's theorem (1931) in mathematical logic and Bell's theorem (1964) in physics share a number of fairly unusual features (for \emph{theorems}):\footnote{\label{God0} In fact there are \emph{two} incompleteness theorems in logic due to  G\"{o}del (see footnotes \ref{God1} and \ref{God2}) and \emph{two} theorems on \qm\  due to Bell  (Brown \& Timpson, 2014; Wiseman, 2014),
 but for reasons to follow in this essay I am mainly interested in the first ones, of both authors, except for a few side remarks.
  }
\begin{itemize}
\item Despite their very considerable technical and conceptual difficulty, both results are extremely famous and have  caught the popular imagination like few others in science.
\item Though welcome in principle--in their teens, many people including the author were  intrigued by books with titles like
\emph{G\"{o}del, Escher Bach: An Eternal Golden Braid} and \emph{The Dancing Wu-Li Masters: An Overview of the New Physics}, both of which appeared in 1979--this imagination has fostered wild claims to the effect that  G\"{o}del proved that the mind cannot be a computer or even that God exists, whilst Bell allegedly showed that reality does not exist. Both theorems (apparently through rather different means) supposedly also supported  the validity of Zen Buddhism.\footnote{\label{God1} See  Franz\'{e}n (2005) for an excellent first introduction to G\"{o}del's theorems, combined with a fair and detailed critique of its abuses, including  overstatements by both amateurs and  experts (a similar guide to the use and abuse of Bell's theorems remains to be written), and Smith (2013) for a possible second go.}
\item However, even among professional mathematicians (logicians excepted) few would be able to correctly state the content of G\"{o}del's theorem when asked on the spot, let alone provide a correct proof, and similarly for Bell's theorem among physicists. 
\item Nonetheless, many professionals will be aware of the general feeling that G\"{o}del in some sense shattered the great mathematician Hilbert's dream of what the foundations of mathematics should look like, whilst there is similar consensus that  Bell dealt a lethal blow to Einstein's physical world view--though ironically, G\"{o}del worked  in the spirit and formalism of Hilbert's proof theory, much as Bell largely \emph{agreed} with Einstein's views about \qm\ and about physics in general.
\item  Both experts and amateurs seem to agree that  G\"{o}del's theorem and Bell's theorem penetrate the very core of the respective disciplines of mathematics and physics.
\end{itemize}
In this light, anyone interested in both of these disciplines
 will want to know what these results have to do with each other, especially since mathematics  underwrites physics (or at least is its language).\footnote{Yanofsky (2013) nicely discusses both theorems in the context of the limits of science and reason.}
At first sight this connection looks remote.
 Roughly speaking:\footnote{Both reformulations are a bit anachronistic and purpose-made. See G\"{o}del (1931) and Bell (1964)!}
 \begin{enumerate}
\item   G\"{o}del proved that any consistent mathematical theory  (formalized as an axiomatic-deductive system in which proofs could in principle be carried out mechanically by a computer) that contains enough arithmetic is incomplete (in that arithmetic sentences $\phv$ exist for which neither $\phv$ nor its negation can be proved).
\item Bell showed that if a deterministic ``hidden variable" theory underneath (and compatible with) \qm\  exists, then this theory cannot be local (in the sense that the hidden state, if known, could be used for superluminal signaling).\end{enumerate}
Both  were triggered by a specific historical context.  G\"{o}del (1931) reflected on the recently developed formalizations of mathematics, of which he specifically mentions the \emph{Principia Mathematica} of Russell and Whitehead and the axioms for set theory proposed earlier by Zermelo, Fraenkel, and von Neumann. Though relegated to a footnote, 
the shadow of \emph{Hilbert's program},
aimed to prove the consistency of mathematics (ultimately based on Cantor's set theory) using absolutely reliable, ``finitist" means, clearly loomed large, too.\footnote{\label{God2} G\"{o}del's \emph{second} incompleteness theorem shows that one example of $\phv$ is the (coded) statement that the consistency of the theory can be proved within the theory. This is often taken to refute Hilbert's program, but even among experts it seems controversial if it really does so. For Hilbert's program and its role in G\"{o}del's theorems see e.g.\ 
Zach (2001), Tait (2005), Sieg (2013), and Tapp (2013).}

Bell, on the other hand, tried to understand if the \emph{de Broglie--Bohm pilot wave theory}, which was meant  to be a deterministic theory of particle motion reproducing all predictions of \qm, \emph{necessarily} had to be non-local: Bell's answer, then, was ``yes."\footnote{\label{Bellfn} Greenstein (2019) is a popular book on the history and interpretation of Bell's work. Scholarly analyses include Redhead (1989), Butterfield (1992), Werner \& Wolf (2001), and the papers cited in footnote \ref{God0}.}

 In turn, the circumstances in which G\"{o}del and Bell operated had a long pedigree in the quest for \emph{certainty in mathematics} 
  and for \emph{determinism in physics}, respectively.\footnote{Some vocal researchers calim that Bell and Einstein were primarily interested in locality and realism, determinism being a secondary (or no) issue, but the historical record is ambiguous; more generally, over 10,000 papers about Bell's theorems show that  Bell can be interpreted in almost equally many ways. But this controversy is a moot point: whatever his own (or Einstein's) intentions, Bell's (1964) theorem puts constraints on possible deterministic underpinnings of \qm, and that is how I take it.
 } The former had even been challenged at least three times:\footnote{For  an overall survey of this theme see Kline (1980). }
   first, by the transition from Euclid's mathematics to Newton's; second, by the set-theoretic paradoxes discovered around 1900 by Russell and others (which ultimately resulted from attempts to make Newton's calculus rigorous by grounding it in analysis, and in turn founding analysis in the real numbers and hence in set theory), and third, by Brouwer's challenge to ``classical" mathematics, which he tried to replace by ``intuitionistic" mathematics (both Hilbert and  G\"{o}del were  influenced by Brouwer, though \emph{contrecoeur}: neither shared his overall philosophy of mathematics). 
   
   In physics (and more generally),
 what Hacking (1990, Chapter 2) calls  the \emph{doctrine of necessity}, which thus far--barring a few exceptions--had pervaded European thought, began to erode in the 19th century, culminating in the invention of \qm\ between 1900--1930 and notably in its probability interpretation as expressed by Born (1926): 
    \begin{quote}\begin{small}
Thus Schr\"{o}dinger's quantum mechanics gives  a  very definite answer to the 
question of  the  outcome  of  a  collision;  however, this does not involve any  causal  relationship.   One obtains 
\emph{no} answer  to  the  question  ``what  is  the  state  after  the  collision,"
but only to the question ``how probable is a specific outcome
 of the collision". (\ldots)
This raises the entire problem of determinism.  From the standpoint of our quantum 
mechanics, there is no quantity that could causally establish the outcome of a collision  
in each individual case;  however,  so far we are not aware of any experimental clue to the effect 
that there are  internal  properties  of  atoms  that enforce some particular outcome.  Should we hope to discover such properties that determine individual outcomes later
(perhaps phases of the internal  atomic  motions)? (\ldots) I myself tend to relinquish determinism in the atomic world.  (Born, 1926, p.\ 866, translation by the present author)
\end{small}\end{quote}
In  a letter to  Born dated December 4, 1926,  Einstein's famously replied that `God does not play dice' (`Jedenfalls bin ich \"{u}berzeugt, da\ss\ \emph{der} nicht w\"{u}rfelt').  Within ten years Einstein saw a link with locality,\footnote{This phase in the history of \qm\ is described by Mehra \&  Rechenberg (2000).} and Bell (1964) and later papers followed up on this.
\section{Randomness and its unprovability}
This precise history has a major impact on my argument,  since it shows that right from the beginning the kind of randomness that Born (probably preceded by Pauli and followed by Bohr, Heisenberg, Jordan, Dirac, von Neumann, and most of the other pioneers of \qm\ except Einstein, de Broglie, and Schr\"{o}dinger) argued for as being produced by \qm, was antipodal to \emph{determinism}.\footnote{See Landsman (2020) for the view that randomness is a  family resemblance (in that it lacks a 
 meaning common to all its applications) with the special feature that its various uses are always defined antipodally.} Thus randomness in \qm\ was identified with \emph{indeterminism}, and hence attempts (like the de Broglie--Bohm pilot wave theory) to undermine the ``Copenhagen" claim of randomness  looked for \emph{deterministic} (and arguably \emph{realistic}) theories underneath \qm. 

Although ``undecidability" may sound a bit like  ``indeterminism", the analogy between the quests for certainty in mathematics and for determinism in physics (and their alleged undermining by G\"{o}del's and Bell's theorems, respectively) may sound rather superficial. To find common ground more effort is needed to bringing these theorems together.\footnote{Also cf.\ e.g.\ Breuer (2001), Calude (2004), Svozil, Calude \& Stay (2005), and  Szangolies (2018).}

First, some of its ``romantic" aspects have to be removed from G\"{o}del's theorem, notably its  reliance on self-reference, although admittedly this \emph{was} the key to both G\"{o}del's original example of an undecidable sentence $\phv$ (which in a  cryptic way expresses its own unprovability) and his proof, in which an axiomatic  theory that includes arithmetic is arithmetized through a numerical encoding scheme so as to be able to ``talk about itself". 
Though later proofs of G\"{o}del's theorem also use numerical encodings of mathematical expressions (such as  symbols, sentences, proofs, and computer programs), this is done in order to make recursion theory (initially a theory of  functions $f:\N\raw\N$) available to a  wider context, rather than to exploit self-reference. Each computably enumerable but uncomputable subset $E\subset \N$ leads to  undecidable statements (very rarely  in mainstream mathematics),\footnote{A subset $E\subset\N$ is  \emph{computably enumerable} (c.e.)  if it is the image of a computable function $f:\N\raw \N$, and \emph{computable} if  its characteristic function $1_E$ is computable, which is true iff both $E$ and $\N\backslash E$ are c.e.} namely those for which the sentence $n\notin E$ is true but unprovable. \emph{Chaitin's (first) incompleteness theorem} (Theorem \ref{Chaitin} in Appendix B), which will play an important role in my reasoning, is an example of this. To understand this theorem and its background we return to the history of 20th century mathematics and physics. 

Hilbert  influenced this history in many ways,\footnote{This is true for physics almost as much as it is (more famously) for mathematics,
since Hilbert played a major role in the mathematization of the two great theories of twentieth century physics, i.e.\  general relativity (Corry, 2004; Renn, 2007) and quantum mechanics  (R\'{e}dei \& St\"{o}ltzner, 2001;  Landsman, 2021).}
of which the sixth problem on his famous list of 23 mathematical
problems from 1900  is particularly relevant here:  this problem concerns the `\emph{Mathematical
Treatment of the Axioms of Physics, especially the theory of probabilities and mechanics}' (Hilbert, 1902). 
This problem  influenced our topic in two initially independent ways, which now come together. First, the problem
inspired von Neumann (1932) to develop his mathematical axiomatization of quantum mechanics, which still forms the
basis of all mathematically rigorous work on this theory. In particular, he initiated the literature on \hv\ theories (see \S\ref{RBT}). 
Second, it led both von Mises and Kolmogorov to their ideas on the mathematical foundations of probability and randomness, initially in opposite ways: whereas von Mises (1919) tried (unsuccessfully) to first axiomatize random sequences of numbers and then extract
probability from these as asymptotic relative frequency, Kolmogorov (1933) successfully axiomatized probability
first and then (unsuccessfully) sought to extract some notion of randomness from this. 

The basic problem (already known to Laplace and perhaps even earlier probabilists) was that, in a 50-50 Bernoulli process for simplicity,
 an apparently ``random" string like
\beq
\sg=001101010111010010100011010111
\eeq
 is as probable as a ``deterministic" string like 
\beq
\sg=111111111111111111111111111111.
\eeq
 In other words, their probabilities say little or nothing about the ``randomness" of individual outcomes. Imposing statistical properties helps but is not enough to guarantee randomness. It is slightly easier to explain this in base 10, to which I therefore switch for a moment. If we call a sequence $x$  \emph{Borel normal} if each possible string $\sg$ in $x$ has relative frequency $10^{-|\sg|}$, where $|\sg|$ is the length of $\sg$
    (so that each digit $0, \ldots, 9$ occurs 10\% of the time,  each block $00$ to $99$ occurs 1\% of the time, etc.), then   \emph{Champernowne's number}
    $$ 0123456789101112131415161718192021222324252629282930 \ldots$$
 can be shown to be Borel normal. The decimal expansion of $\pi$ is conjectured to be Borel normal, too (and has been empirically verified to be so in billions of decimals), but these numbers are hardly random: they are computable, which is one version of ``deterministic".
 
 Any sound definition of randomness (for binary strings or sequences) has to navigate between Scylla and Charybdis: if the definition is too weak (such as Borel normality), counterexamples will undermine it (such as Champernowne's number), but if it is too strong (such as being lawless, like Brouwer's choice sequences), it will not hold almost surely in a 50-50 Bernoulli process
 (Moschovakis, 2016).
As an example of such sound navigation, Solomonoff, Kolmogorov, Martin-L\"{o}f, Chaitin, Levin, Solovay, Schnorr, and others developed the \emph{algorithmic theory of randomness} (Li \& Vit\'{a}nyi, 2008). 
 The basic idea is that a string or sequence is random iff its shortest description is the sequence itself, but the notion of a description has to made precise to avoid \emph{Berry's paradox}:
\begin{quote}\begin{small}
The Berry number is the smallest positive integer that cannot be described in less than eighteen words.
\end{small}\end{quote}
The paradox, then, is that on the one hand this number must exist, since only finitely many integers can be described in less than eighteen words and hence the set of such numbers must have a lower bound, while on the other hand Berry's number cannot exists by its own definition.\footnote{This is one of innumerable paradoxes of natural language, which
leads to an incompleteness theorem once the notion of a description has been appropriately formalized in mathematics, much as G\"{o}del's first incompleteness theorem turns the the liar's paradox into a theorem. }
In the case at hand, the notion of a \emph{description} is sharpened by asking it to be \emph{computable}, so that, roughly speaking (see appendix B for technical details), we call a (finite) binary string $\sg$ \emph{(Kolomogorov) random} 
if the  length of the shortest computer program generating $\sg$ is at least as long as $\sg$ itself, and call 
 an (infinite) binary sequence $x$ \emph{(Levin--Chaitin) random} or \emph{1-random} if its (sufficiently long)  finite truncations are Kolomogorov random. At last, for finite strings $\sg$
 \emph{Chaitin's (first) incompleteness theorem} states that although countably many strings $\sg$ \emph{are} random, this can be \emph{proved} only for finitely many of these, whereas for infinite sequences $x$ his (second) incompleteness theorem says that if such a sequence is random, only finitely many of its digits can be computed (see Theorems \ref{Chaitin} and \ref{Klaas} for precise statements). Thus
  \emph{randomness is elusive}. 
\section{Rethinking Bell's theorem}\label{RBT}
In order to locate Bell's (1964) theorem in the literature on \qm\ and (in)determinism, I recall that Hilbert's sixth problem inspired both the work of von Mises and Kolmogorov that eventually gave rise to the algorithmic theory of randomness, \emph{and} (Hilbert's postdoc) von Neumann's work on the mathematical foundations of \qm, culminating in his book (von Neumann, 1932). One of his results was that there can be no nonzero function $\lm: H_n(\C)\raw \R$ (where $H_n(\C)$ is the space of hermitian $n\x n$ matrices, seen as the observables of a quantum-mechanical $n$-level system) that is:
\begin{enumerate}
\item \emph{dispersion-free} (i.e.\ $\lm(a^2)=\lm(a)^2$ for each $a\in H_n(\C)$);
\item \emph{linear} (i.e.\ $\lm(sa+tb)=s\lm(a)+t\lm(b)$ for all $s,t\in\R$ and $a,b\in H_n(\C)$).
\end{enumerate}
Unfortunately, von Neumann interpreted this correct, non-circular, and interesting result as a proof that \qm\ is complete in the sense that there can be no hidden variables in the sense of Born (1926), i.e.\ `properties that determine individual outcomes'. The reason this does not follow is twofold.\footnote{See also
 Bub (2011), Dieks (2016), and forthcoming work by Chris Mitsch for balanced accounts.} First, the proof relies on a tacit assumption that later came to be called \emph{non-contextuality}, namely that the value $\lm(a)$ of some observable $a$ \emph{only depends on $a$}, whereas  measurement ideology \`{a} la Bohr (1935) suggests that it may depend on a \emph{measurement context}, formalized  as a further set of observables commuting with $a$ (unless $a$ is maximal such a set is far from unique).\footnote{
 The idea of contextuality was first formulated by Grete Hermann (Crull \& Bacciagaluppi, 2016). 
 } Second, though natural, the linearity assumption is very strong and excludes even eigenvalues of $a$. 

This second point was remedied by the \emph{Kochen--Specker theorem},\footnote{See Kochen \& Specker (1967). Ironically, his followers attribute this theorem to Bell (1966), although the result is just a technical sharpening of von Neumann's result they so vehemently ridicule. For a deep philosophical analysis of the Kochen--Specker theorem, as well as of Bell's theorems,  see Redhead (1989). } who weakened von Neumann's linearity assumption to \emph{linearity on commuting observables}, which at least incorporates eigenvalues and is even found so appealing that the Kochen--Specker is generally taken to exclude non-contextual \hv\ theories. See also Appendix C. 

The final step in the series of attempts, initiated by von Neumann, to exclude hidden variables by showing that  subject to reasonable assumptions the corresponding value attributions cannot exist even independently of any statistical considerations, is the so-called \emph{free will theorem}.\footnote{See appendix C.
This theorem is originally due to Heywood \& Redhead (1983), with follow-ups by Stairs (1983),
 Brown \& Svetlichny (1990),  and Clifton (1993), but
it was named and made famous by Conway \& Kochen (2009), whose main contribution was an emphasis on free will  (Landsman, 2017, Chapter 6). } In the wake of the renowned ``{\sc epr}'' paper (Einstein, Podolsky and Rosen, 1935) the setting has now become bipartite (i.e.\ Alice and Bob who are spacelike separated each perform experiments on a correlated state) and the \emph{non-contextuality} assumption is weakened to \emph{local contextuality}: the outcomes of Alice's measurements are independent of any  choice of measurements Bob might perform, and \emph{vice versa}.\footnote{Since Alice and Bob are spacelike separated their observables commute (Einstein locality).} Thus her value attributions $\lm(a| \mathrm{context})$ may well be contextual, \emph{as long as the observables commuting with the one she measures (i.e.\ $a$), which form a context to $a$, are local to her. }

A second line of research, which goes back at least to de Broglie (1928), was influentially taken up by Bohm (1952),
and most recently includes 't Hooft (2016), assumes the possibility of non-contextual value attributions and tries to make these compatible with the Born rule of \qm.  Bell (1964) was primarily concerned with such theories, asking himself if a deterministic theory like Bohm's was necessarily non-local.

\noindent In Bell's analysis, which takes place in the bipartite ({\sc epr}) setting, the quantum-mechanical probabilities are obtained by formally averaging over the set of hidden variables, i.e.,
\begin{equation}
P_{\psi}(F=x,G=y\mid A=a, B=b)=\int_{\Lm} d\mu_{\psi}(\lm)\, P_{\lm}(F=x,G=y\mid A=a, B=b).\label{Bell}
\end{equation}
Here $\ps$ is some (explicitly identified) quantum state of a correlated pair of (typically) 2-level quantum systems (which may be either optical, where the degree of freedom is helicity, or massive, where the degree of freedom is spin),
$F$ is an observable measured by Alice defined by her choice of setting $a$, likewise $G$ for Bob defined by his setting $b$, with possible outcomes $x\in\ul{2}=\{0,1\}$, likewise $y\in\ul{2}$ for Bob; the left-hand side is the Born probability for the outcome $(x,y)$ if the correlated system has been prepared in the state $\psi$; the expression $P_{\lm}(\cdots)$ on the right-hand side is the probability of the outcome $(x,y)$ if the unknown hidden variable or state equals $\lm$,  and finally, $\mu_{\psi}$ is some probability measure on the space $\Lm$ of hidden states supposedly provided by the theory for each
state  $\ps$. 

We now say that the \hv\ theory supplying the above quantities is:
\begin{itemize}
\item  \emph{deterministic} if the  probabilities $P_{\lm}(F=x,G=y\mid A=a, B=b)$ equal 0 or 1;
\item  \emph{locally contextual} if the expression
\begin{equation}
P_{\lm}(F=x\mid A=a, B=b)= \sum_{y=0,1} P_{\lm}(F=x,G=y\mid A=a, B=b);\label{A}
\end{equation}
 is independent of $b$, whilst the corresponding expression
\begin{equation}
P_{\lm}(G=y\mid A=a, B=b)= \sum_{x=0,1} P_{\lm}(F=x,G=y\mid A=a, B=b), \label{B}
\end{equation}
 is independent of $a$. That is, the probabilities of Alice's outcomes are independent of Bob's settings, and \emph{vice versa}. This locality property seems very reasonable and in fact it follows from special relativity, for if Bob chooses his settings just before his measurement, there is a frame of reference in which Alice measures before Bob has chosen his settings,  and \emph{vice versa}. In turn, this is equivalent to the property that even if she knew the value of $\lm$, Alice could not signal  to Bob, and \emph{vice versa}.\footnote{In \qm\  the left-hand side of \er{Bell} satisfies this locality condition for any state $\psi$.}
\end{itemize}
Bell proved that  a \hv\ theory cannot satisfy \er{Bell} and be both deterministic and locally contextual (which explained why Bohm's theory had to be non-local). 
Making his  \emph{tacit}
assumption that  experimental settings can be ``freely" chosen \emph{explicit}, we obtain:\footnote{See Landsman (2017), \S 6.5 for details, or Appendix C below for a summary.}
\begin{theorem}\label{BT}
The conjunction of the following properties is inconsistent:
\begin{enumerate}
\item \emph{determinism};
\item  \emph{quantum mechanics}, i.e.\  the Born rule for $P_{\psi}(F=x,G=y\mid A=a, B=b)$;
\item \emph{local contextuality};
\item  \emph{free choice}, i.e.\ (statistical) independence of the measurement settings $a$ and $b$ from each other and from the \hv\ $\lm$ (given the probability measure $\mu_{\psi}$). 
\end{enumerate}
\end{theorem}
\section{Are deterministic \hv\ theories deterministic?}
Although the assumptions have a slightly different meaning, the free will theorem leads to the same result as Bell's theorem (see Appendix C), so that the (no) \hv\ tradition initiated by von Neumann, which culminates in the former, 
coalesces with the (positive) \hv\ tradition going back to de Broglie, shown its place by the latter.
Thusly there are the obvious four (minimal) ways out of the contradiction in Theorem \ref{BT}:
 \begin{itemize}
\item Copenhagen (i.e.\ mainstream) \qm\ rejects determinism;
\item Valentini (2019) rejects the Born rule and hence {\sc qm} (see the end of \S\ref{S4} below); 
\item Bohmians reject  local contextuality;\footnote{There is a subtle difference between Bohmian mechanics as reviewed by e.g.\ Goldstein (2017), and de Broglie's original pilot wave theory 
(Valentini, 2019). This difference is immaterial for my discussion.}
\item 't Hooft (2016) rejects free choice.  
\end{itemize}
We focus on the last two options, so that determinism and quantum mechanics (i.e.\ the Born rule) are kept.
 In both cases the Born rule is  recovered by averaging the hidden variable with respect to 
 a probability measure $\mu_{\ps}$ on the space of hidden variables, given some (pure) quantum state $\ps$. The difference is that in Bohmian mechanics the total state (which consists of the hidden configuration plus the ``pilot wave'' $\ps$) determines the measurement outcomes \emph{given the settings}, whereas in  't Hooft's theory 
 the hidden variable determines the outcomes as well as the settings.\footnote{In Bohmian mechanics, 
 the hidden state $q\in Q$ just pertains to the  particles undergoing measurement, whilst  the settings $a$ are supposed to be ``freely chosen" for each measurement (and in particular are independent of $q$).
   The outcome is then fixed by $a$ and $q$.  In  't Hooft's  theory, the hidden state $x\in X$ of ``the world" determines the settings as well as the outcomes. Beyond the issue raised in the main text, Bohmians (but not 't Hooft!) therefore have an additional problem, namely the origin of 
 the  settings (which are simply left out of the theory). This weakens their case for determinism even further.}   More specifically:
 \begin{itemize}
\item  In Bohmian mechanics the hidden variable is position $q$, and $d\mu_{\ps}(q)=|\psi(q)|^2 dq$ is  the Born probability for outcome $q$ with respect to the expansion $|\psi\ra=\int dq\, \psi(q) |q\ra$.
\item In 't Hooft's theory the hidden variable  is a basis vector $|m\ra$ in some separable Hilbert space $H$ ($m\in\N$), and once again the measure $\mu_{\ps}(m)=|c_{m}|^2$  is given by the Born probability for outcome $m$ with respect to the expansion $|\psi\ra=\sum_m c_m|m\ra$. 
\end{itemize}
Thus the hidden variables (i.e.\ $q\in Q$ and $m\in\N$, respectively) have familiar quantum-mechanical interpretations  and also their compatibility measures are precisely the Born measures for the quantum state $\psi$. In this light, we may ask to what extent these \hv\ theories are truly deterministic, as their adherents claim them to be. 
Since the argument does not rely on entanglement and hence on a bipartite experiment, we may as well work with a quantum coin toss. The settings, possible contexts, and quantum state of the experiments are then fixed, 
 so that only the hidden state $\lm$ may change and hence we may treat Bohmian mechanics and 't Hooft's theory on the same footing.
 Idealizing to an infinite run, one has an outcome sequence $x:\N\raw\ul{2}$.
 Standard (Copenhagen) \qm\ refuses to say anything about its origin, but nonetheless it does make very specific predictions about $x$. The basis of these predictions is the following theorem, whose notation and proof are explained in Appendix A. 
 One may think of a  fair quantum coin, in which $\sg(a)=\ul{2}=\{0,1\}$ and $\mu_a(0)=\mu_a(1)=1/2$, and which  \emph{probabilistically}
 is indistinguishable from a fair classical coin (which in my view cannot  exist, cf.\  \S\ref{S4}).
 \begin{theorem}\label{ET}
The following procedures for repeated identical independent measurements are equivalent (in giving the same possible outcome sequences with the same probabilities):
 \begin{enumerate}
 \item Quantum mechanics is applied to the whole run, described  as a single quantum-mechanical experiment with a single classically recorded outcome sequence;
\item Quantum mechanics is  applied to single experiments (with classically recorded outcomes), upon which classical probability theory takes over to combine these.
\end{enumerate}
Either way, the (purely theoretical) Born probability $\mu_a$ for single outcomes induces the infinite Bernoulli process probability 
$\mu_a^{\infty}$ on the space $\sg(a)^{\N}$ of infinite outcome sequences.
\end{theorem}
 Theorem \ref{PML} in Appendix B then implies:
  \begin{corollary}\label{keycor}
  With respect to the ``fair'' probability measure $P^{\infty}$ on $\ul{2}^{\N}$ almost every outcome sequence $x$ of an infinitely often repeated fair quantum coin flip is
  1-random.
 \end{corollary}
In
 \hv\ theories, on the other hand,  $x$ factors through $\Lm$, that is,\footnote{The function $g$ incorporates all details of the experiment that may affect the outcome (like the setting,  context, and quantum state) except the hidden variable $\lm$ (which it \emph{specifies}). It has nothing to do with non-contextual value assignments on the set of quantum-mechanical observables (which do not exist). } there are functions $h:\N\raw\Lm$ and $g:\Lm\raw\ul{2}$ such that $x=g\circ h$. Hidden variable theories do provide $g$, i.e.\ describe the outcome of any experiment given the value of the hidden variable $\lm\in\Lm$. 
  However, what about $h$, that is, the specification of the value of the \hv\ $\lm$ in each run of the experiment?
There are just the following two scenarios:
\begin{enumerate}
\item  \emph{The function $h$ is provided by the hidden variable theory.} In that case, since the theory is supposed to be deterministic,  $h$ explicitly gives the values $\lm_n=h(n)$ for each $n\in\N$ (i.e.\ experiment no.\ $n$ in the run). 
Since $g$ is also given, this means that $x$ is given by the theory. By Theorem \ref{Klaas} (i.e.\ Chaitin's second incompleteness theorem), the outcome sequence cannot be 1-random, against Corollary \ref{keycor}.
 \item  \emph{The function $h$ is not provided by the hidden variable theory.}  In that case, the theory fails to determine the outcome of any specific experiment and just provides averages of outcomes. My conclusion would be that, except for some kind of a ``story'', nothing has been gained over \qm, but \hv\ theorists argue that their theories cannot be expected to provide initial conditions (for experiments), and claim that the randomness in measurement outcomes originates 
  in the randomness of the initial conditions of the experiment.\footnote{The Bohmians are  divided on the origin of their compatibility measure, referred to in this context as the  \emph{quantum equilibrium distribution}, cf.\ 
D\"{u}rr, Goldstein, \& Zanghi (1992) against Valentini (2019).  The origin of $\mu_{\psi}$  is not my concern, which is the need to randomly sample it and the justification for doing so.
}  
But then the question arises what else provides these conditions, and hence our function $h$. The point here is that in order to recover the predictions of \qm\ as meant in  Corollary \ref{keycor},
  the function $h$ must sample the Born measure (in its guise of the compatibility measure $\mu_{\psi}$ on $\Lm$), in the sense of  ``randomly" picking elements from  $\Lm$, distributed according to  $\mu_{\psi}$, cf.\ \er{nuas5}.
   This, in turn, should guarantee that the sequences $x=g\circ h$
   mimic fair coin flips. Since $g$ is  supposed to be given, this implies that the randomness properties of $x$ must entirely originate in $h$.  This  origin cannot be deterministic, since in that case we are back to the contradictory scenario 1 above. Hence $h$ must come from some unknown external random process in nature that our hidden variable theories invoke as a kind of an oracle.
   In my view the need for such a random oracle undermines their purpose and makes them self-defeating.  Every way you look at this you lose!
   \end{enumerate}
    \section{Conclusion and discussion}\label{S4}
    We may summarize the discussion in the previous section as follows:\footnote{In stating the second condition I have taken $\sg(a)=\{0,1\}$ with 50-50 Born probabilities, but this can be generalized to other spectra and probability measures.
    See  Downey \& Hirschfeldt (2010), \S 6.12.} 
    \begin{theorem}\label{LT}
For any \hv\ theory $T$ the following properties are incompatible:
\begin{enumerate}
\item \emph{Determinism:} $T$ states the outcome of the measurement of any observable $a$  \emph{given the value $\lm\in\Lm$
of the \hv} via a function  $g:\Lm\raw \sg(a)$  \emph{and} provides these values for each 
 experiment; for an infinite run this is done via some function $h:\N\raw\Lm$, so that $T$ provides the outcome sequence $x:\N\raw\sg(a)$ through $x=g\circ h$.
\item  \emph{Born rule:}  Outcome sequences are almost surely 1-random.
  (cf.\ Corollary \ref{keycor}). 
\end{enumerate}
\end{theorem}
The proof is short. According to the first clause $T$ states the entire outcome sequence $x$.  By Chaitin's incompleteness theorem \ref{Klaas} this is incompatible with the second clause. \hfill $\Box$
\smallskip

\noindent
 In order to understand Theorem  \ref{LT} and its proof
it may be helpful to note that in classical coin tossing the role of the hidden state is also played by  the initial conditions (cf.\ Diaconis \& Skyrms, 2018 Chapter 1, Appendix 2). The 50-50 chances (allegedly) making the coin fair are obtained by averaging over the initial conditions, i.e., by sampling. By the same arguments, this sampling cannot be deterministic, i.e.\ given by a function like $h$, for otherwise the outcome sequences appropriate to a fair coin would not obtain: it must be done in a genuinely random way and hence by appeal to an external random process. This is impossible classically, so that--unless they have a quantum-mechanical seed--\emph{fair classical coins do not exist}, as  confirmed by Diaconis \& Skyrms (2018, Chapter 1).

I conclude that deterministic \hv\ theories compatible with \qm\ do not exist. The reason that 
Bell's (1964) theorem and the free will theorem leave two loopholes for determinism (i.e.\ local contextuality  and no free choice) is that their compatibility condition with \qm\ is  stated too weakly: 
the theory is only required to reproduce certain single-case (Born) probabilities, as opposed to the properties of typical outcome sequences (from which the said probabilities are extracted as long-run frequencies). This reason this approach is still partly successful lies in the clever use of entangled states. If one rejects the second requirement on determinism in Theorem \ref{LT}, Bell's theorem and the free will theorem still provide useful constraints on deterministic \hv\ theories, but as shown in the previous section such a rejection necessitates an appeal to an unknown random process and hence seems self-defeating. 

Let us now consider the role of the idealization to infinite outcome sequences and see what happens if the experimental runs are finite.\footnote{In other words, we examine whether Earman's principle is satisfied, cf.\ footnote \ref{Ear}.}
Once again, via Theorem \ref{ET}  the Born rule predicts that outcome strings will be Kolmogorov random with high probability. Any deterministic theory (in the sense of Theorem \ref{LT}) provides an explicit description (say in ZFC) of the outcomes, whose randomness would be provable from this description. But this is precluded by Chaitin's first incompleteness theorem (i.e.\ Theorem \ref{Chaitin}), now in the role played by his second incompleteness theorem in the infinite case.\footnote{To make this argument completely rigorous one would need to define what a ``description'' provided by a deterministic theory means logically. There is a logical characterization of deterministic theories (Montague, 1974), and there
are some arguments to the effect that the evolution laws in deterministic theories should be computable, cf.
Earman (1986), Chapter 11, and Pour-El \& Richards (2016), \emph{passim}, but this literature makes no direct reference to output strings or sequences of the kind we analyze and in any case the identification of ``deterministic'' with ``computable'' is obscure even in situations where the latter concept is well defined. For example, if we stipulate that $h:\N\raw\Lm$ is computable (and likewise $g:\Lm\raw\ul{2}$) then the above appeal to  Chaitin's first incompleteness theorem is not even necessary, but this seems too easy. A  somewhat circular solution, proposed by Scriven (1957), is to simply say that $T$ is deterministic iff the output strings or sequences it describes are not random, but this begs for a more explicit characterization. One might naively expect such a characterization to come from the \emph{arithmetical hierarchy} (found in any book on computability): if, as before, we identify $\ul{2}^{\N}$ with the power set $P(\N)$ of $\N$, then  $S\subset\N$ is called \emph{arithmetical} if there is a formula $\ps(x)$ in PA (Peano Arithmetic) such that $n\in S$ iff $\N\vDash \psi(n)$,  that is, $\psi(n)$ is true in the usual sense. We may then classify the arithmetical subsets through the logical form of $\ps$, assumed in prenex normal form (i.e., all quantifiers have been moved to the left): $S$ is in $\Sg_0^0=\Pi_0^0$ iff $\psi$ has no quantifiers or only bounded quantifiers (in which case $S$ is computable), and then recursively $S\in\Sg^0_{n+1}$ iff $\psi(x)=\exists_y\phv(x,y)$ with $\phv\in \Pi_n^0$, and  $\phv\in \Pi_{n+1}^0$ iff $\psi(x)=\forall_y\phv(x,y)$ with $\phv\in \Sg_n^0$. Here any singly quantified
expression $\exists_y\phv(x,y)$ may  be replaced by  $\exists{y_1}\cdots \exists{y_k}\phv(x,y_1, \ldots, y_k)$ and likewise for $\forall_y$.  By convention  $\Sg^0_n\subset\Sg^0_{n+1}$ and 
$\Pi^0_n\subset\Pi^0_{n+1}$, and  $\Dl^0_n:=\Sg_n^0\cap\Pi_n^0$. Since in classical logic $\forall_y\phv(x,y)$ is equivalent to $\neg\exists_y\neg\phv(x,y)$, it follows that $\Pi_n^0$ sets are the complements of $\Sg_n^0$ sets. One would then like to locate deterministic theories somewhere in this hierarchy, preferably above the computable $\Dl_0^0$. The idea of a \hv\ (namely $y$) suggests $\Sg_1^0$ and closure under complementation (it would be crazy if some deterministic theory prefers ones over zeros) then leads to $\Dl_1^0$, but this equals $\Dl_0^0$. The next level  $\Dl_2^0$ is impossible since this already contains 1-random sets like Chaitin's $\Om$. Hence more research is needed. } \hfill $\Box$

Nonetheless, although their incompatibility with \qm\ has now been established, it will be hard to disprove deterministic \hv\ theories from  experimental data. Let us look at the proof of Bell's theorem for inspiration as to what such a (dis)proof should look like. In the context of the {\sc epr}--Bohm experiment local deterministic \hv\ theories  predict correlations that \emph{satisfy} the Bell inequalities,\footnote{For Bell's proof it is irrelevant whether or not some \hv\ is able to sample the compatibility measure, since the Bell inequalities follow from pointwise bounds, cf.\   Landsman (2017), eq.\ (6.119).}
  whereas on suitable settings  \qm\ predicts (and experiment shows) that typical outcome sequences \emph{violate} these inequalities. 
Now a disproof of some deterministic  \hv\ theory $T$ cannot perhaps be expected to show that \emph{all}  quantum-mechanical outcome sequences violate the predictions of  the \hv\ theory (indeed they do not, albeit with low probability), but it should identify at least a sufficiently large number of  typical (i.e.\ random) sequences. However, even in the finite case this identification is impossible by Theorem \ref{Chaitin}, so that the false predictions of $T$ cannot really be confronted with the correct predictions of \qm.
 Thus the unprovability of their
falsehood condemns deterministic \hv\ theories, and perhaps even determinism as a whole, to a zombie-like existence in a twilight zone comparable with the Dutch situation around selling soft drugs: although this is forbidden by law, it is (officially) not prosecuted. 
 
The situation would change drastically  if deterministic \hv\ theories gave up their compatibility with the Born rule (on which my entire reasoning is based), as for example Valentini (2019) has argued in case of the de Broglie--Bohm pilot wave theory. For it is this compatibility requirement that kills such theories, which could leave zombie-dom if only they were brave enough to challenge the Born rule. This might open the door to superluminal signaling and worse, but on the other hand the possibility of violating the Born rule would also provide a new context for deriving it, e.g.\ as a dynamical equilibrium condition (as may be the case for the  Broglie--Bohm theory, if Valentini is right). 

I would personally expect that the Born rule is emergent from some lower-level theory, which equally well suggests that it is  valid in some limit only, rather than absolutely.
\smallskip

\noindent \begin{footnotesize}The author is grateful to Jacob Barandes, Jeremy Butterfield, Cristian Calude, Erik Curiel, John Earman, Bas Terwijn, and Noson Yanofsky, as well as to  members of seminar audiences and especially readers of the first version of this essay on the FQXi website for very helpful comments and corrections. He is even more grateful to the late Michael Redhead, for his exemplary approach to the foundations of physics. 
\end{footnotesize}
\newpage
 \appendix
 \section{The Born rule}\label{BR}
The Born measure is a  probability measure $\mu_a$ on the spectrum $\sg(a)$ of a (bounded) self-adjoint operator $a$ on some \Hs\ $H$, defined as follows by any state $\om$ on $B(H)$:\footnote{
Here a state $\om$ is a positive normalized linear functional on $B(H)$, as in the \ca ic approach to \qm\ (Haag, 1992; Landsman, 2017). One may think of expectation values 
$\om(a)=\Tr(\rh a)$, where $\rh$ is a density operator on $H$, with the special case $\om(a)=\la \ps, a\ps\ra$, where $\ps\in H$ is a unit vector. For a proof of Theorem \ref{defBornmu} see e.g.\ Landsman (2017), \S 4.1, Corollary 4.4.
}
 \begin{theorem}\label{defBornmu}
 Let $H$ be a \Hs, let $a^*=a\in B(H)$, and let $\om$ be a state on $B(H)$. 
  There exists a unique probability measure 
 $\mu_a$ on the spectrum $\sg(a)$ of $a$  such that 
 \begin{equation}
\om(f(a))=\int_{\sg(a)} d\mu_a(\lm)\, f(\lm), \:\: \mathrm{ for\: all }\:\: f \in C(\sg(a)).\label{BornfromGelfand}
\end{equation}
  \end{theorem}
 The Born measure is a mathematical construction; what is its relationship to experiment? This relationship must be the source of the (alleged) randomness of \qm, for the Schr\"{o}dinger equation is deterministic. We 
start by postulating, as usual, that $\mu_a(\Dl)$ is the (single case) ``probability'' that measurement of the observable $a$ in the state $\om$ (which jointly give rise to the pertinent Born measure $\mu_a$) gives a result $\lm\in\Dl\subset\sg(a)$.  Here  I identify single-case ``probabilities'' with numbers (consistent with the probability calculus) provided by \emph{theory}, upon which long-run frequencies provide \emph{empirical evidence} for the theory in question,
 but do not \emph{define} probabilities. The Born measure is a case in point: these probabilities are \emph{theoretically given},  but have to be \emph{empirically verified} by long runs of
 independent experiments. In other words, by the results reviewed below such experiments provide numbers whose role it is to test the Born rule as a hypothesis. This is justified by the following \hi{sampling theorem} (strong law of large numbers): for any (measurable) subset $\Dl\subset \sg(a)$ and any sequence $(x_n)\in \sg(a)^{\N}$ we have
  $\mu_a^{\infty}$-almost surely:
 \begin{equation}
\lim_{N\raw\infty} \frac{1}{N}(1_{\Dl}(x_1)+\cdots + 1_{\Dl}(x_N))= \mu_a(\Dl).  \label{nuas5}
\end{equation}
  \emph{Proof of Theorem \ref{ET}}.
Let $a=a^*\in B(H)$, where $H$ is a \Hs\  and $B(H)$ is the algebra of all bounded operators on $H$, and let $\sg(a)$ be the spectrum of $a$. For simplicity (and since this is enough for our applications, where $H=\C^2$) I assume $\dim(H)<\infty$, so that $\sg(a)$ simply consists of the eigenvalues $\lm_i$ of $a$ (which may be degenerate). Let us first consider a \emph{finite} number $N$ of  identical measurements of $a$ (a ``run''). 
The first option in the theorem corresponds to a simultaneous measurement of the commuting operators 
\begin{align}
a_1&=a\ot 1_H\ot\cdots\ot 1_H; \label{a1} \\
& \cdots \nn \\
a_N &=1_H\ot\cdots\ot 1_H\ot a, \label{aN}
\end{align}
all defined on the $N$-fold tensor product $H^N\equiv H^{\ot N}$ of $H$ with itself.\footnote{This can even be replaced by a single measurement, see Landsman (2017), Corollary A.20.}  To put this in a broader perspective,  consider \emph{any} set $(a_1, \ldots, a_N)\equiv\ul{a}$ of commuting operators on \emph{any} \Hs\ $K$ (of which \er{a1} - \er{aN} is obviously a special case with $K=H^N$).
These operators have a \emph{joint spectrum} $\sg(\ul{a})$, whose elements are the \emph{joint eigenvalues} $\ul{\lm}=(\lm_1, \ldots, \lm_N)$, defined by the property that there exists a nonzero joint eigenvector $\ps\in K$ such that $a_i\ps=\lm_i\ps$ for all $i=1, \ldots, N$; clearly, 
\begin{equation}
\sg(\ul{a})=\{\ul{\lm}\in
\sg(a_1)\x\cdots\x\sg(a_n)\mid e_{\ul{\lm}}\equiv e^{(1)}_{\lm_1}\cdots e^{(n)}_{\lm_n} \neq 0\} \subseteq \sg(a_1)\x\cdots\x\sg(a_N), \label{21}
\end{equation}
where $e^{(i)}_{\lm_i}$ is the spectral projection of $a_i$ on the eigenspace for the eigenvalue $\lm_i\in\sg(a_i)$.
Von Neumann's Born rule for the probability of finding $\ul{\lm}\in\sg(\ul{a})$ then simply reads 
\begin{equation}
p_{\ul{a}}(\ul{\lm})=\om(e_{\ul{\lm}}),\label{paQn}
\end{equation}
 where $\om$ is the state on $B(K)$ with respect to which the Born probability is defined.\footnote{The uses of states themselves may be justified by Gleason's theorem (Landsman, 2017, \S\S 2,7, 4.4).}
 If $\dim(K)<\infty$, as I assume, we always have $\om(a)=\Tr(\rh a)$ for some density operator $\rh$, and for a  general \Hs\ $K$ this is the case iff the state $\om$ is normal on $B(K)$. For (normal) pure states we have $\rh=|\psi\ra\la\psi|$ for some unit vector $\ps\in K$, in which case
 \begin{equation}
p_{\ul{a}}(\ul{\lm})=\la\ps,e_{\ul{\lm}}\psi\ra.\label{paQnp}
\end{equation}
The  Born rule \er{paQn} is similar to the single-operator case  (Landsman, 2017,
 \S 4.1):\footnote{The  Born rule for  commuting operators follows from the single operator case  (Landsman, 2017, \S 2.5).}
the continuous functional calculus gives a Gelfand isomorphism of commutative C*-algebras
\begin{equation}
C^*(\ul{a},1_K)\cong C(\sg(\ul{a})),
\end{equation}
 under which the restriction of the state $\om$,  originally defined on $B(K)$, to its commutative C*-subalgebra $C^*(\ul{a})$ defines a probability measure $\mu_{\ul{a}}$ on the joint spectrum $\sg(\ul{a})$ via the Riesz isomorphism. This is  the Born measure, whose probabilities are given by \er{paQn}.
For the case  \er{a1} - \er{aN} we have equality in \er{21}; since in that case $\sg(a_i)=\sg(a)$,  we obtain
\begin{equation}
\sg(\ul{a})=\sg(a)^N,
\end{equation}
and therefore, for all $\lm_i\in\sg(a)$ and states $\om$ on $B(H^N)$, 
 the Born rule \er{paQn} becomes 
 \begin{equation}
p_{\ul{a}}(\lm_1, \ldots, \lm_N)=\om(e_{\lm_1}\ot\cdots\ot e_{\lm_N}).
\end{equation}
Now take a state $\om_1$ on $B(H)$. Reflecting the idea that $\om$ is the state on $B(H^N)$ in which $N$ independent measurements of $a\in B(H)$ in the state $\om_1$ are carried out,
choose
 \beq
 \om=\om_1^N,
 \eeq
  the state on $B(H^N)$ defined by linear extension of its action on elementary tensors:
\beq
\om_1^N(b_1\ot\cdots\ot b_n)=\om_1(b_1) \cdots \om_N(b_N).
\eeq
It follows that
\begin{equation}
\om^N(e_{\lm_1}\ot\cdots\ot e_{\lm_N})=\om_1(e_{\lm_1})\cdots\om_1(e_{\lm_N})=p_a(\lm_1)\cdots p_a(\lm_N),
\end{equation}
so that the joint probability of the outcome $(\lm_1, \ldots, \lm_N)\in\sg(\ul{a})$ is simply
\begin{equation}
p_{\vec{a}}(\lm_1, \ldots, \lm_N)=p_a(\lm_1)\cdots p_a(\lm_N).
\end{equation}
Since these are precisely the probabilities for option 2 (i.e.\ the Bernoulli process), i.e.,
 \beq
 \mu_{\ul{a}}=\mu_a^N, \label{mulaN}
 \eeq
 this
 proves the claim for $N<\infty$.  
 To describe the limit $N\raw\infty$, let $B$ be any C*-algebra with unit $1_B$;  below I take $B=B(H)$, $B=C^*(a,1_H)$,  or $B=C(\sg(a))$. 
 We now take
 \beq
 A_N=B^{\ot N},
 \eeq
  the $N$-fold tensor product 
  of $B$ with itself.\footnote{If $B$ is infinite-dimensional, for technical reasons  the so-called \emph{projective} tensor product should be used.} The  special cases above may  be rewritten as
\begin{align}
B(H)^{\ot N}&\cong  B(H^N);\label{BHN}\\
C^*(a,1_H)^{\ot N}&\cong C^*(a_1, \ldots, a_N,1_{H^N});\label{StaN} \\
C(\sg(a))^{\ot N}&\cong  C(\sg(a) \x\cdots\x\sg(a)), \label{sgN}
\end{align}
 with $N$ copies of $H$ and $\sg(a)$, respectively, and in \er{StaN} the $a_i$ are given by \er{a1} - \er{aN}.
 We may then wonder if these algebras have a limit as $N\raw\infty$. They do, but it is not unique and depends on the choice of observables, that is, of the infinite sequences $\mathbf{a}=(a_1,a_2, \ldots)$, with $a_N\in A_N$, that are supposed to have a limit. One possibility is to take sequences $\mathbf{a}$ for which  there exists $M\in\N$ and $a_M\in A_M$ such that for each $N\geq M$,
\beq
a_N=a_M \ot 1_B\cdots\ot 1_B, \label{anam}
\eeq
 with $N-M$ copies of  $1_B$. On that choice, one obtains the infinite tensor product $B^{\ot\infty}$, see Landsman (2017), \S C.14. The limit of \er{BHN} in this sense is $B(H^{\ot\infty})$, where $H^{\ot\infty}$ is von Neumann's `complete' infinite tensor product of \Hs s,\footnote{See Landsman (2017), \S 8.4 for this approach. The details are unnecessary here.}  
in which $C^*(a,1_H)^{\ot\infty}$ is the C*-algebra generated by $(a_1,a_2, \ldots)$ and the unit on $H^{\ot\infty}$.
 The limit of \er{sgN} is 
 \begin{equation}
C(\sg(a))^{\ot \infty}\cong C(\sg(a)^{\N}), 
\end{equation}
where $\sg(a)^{\N}$, which we previously  saw as a measure space (as a special case of $X^{\N}$ for general compact Hausdorff spaces $X$), is now seen as a topological space with the product  topology, in which it is compact.\footnote{Cf.\ Tychonoff's theorem. The associated Borel structure is the one defined by the cylinder sets.} As in the finite case, we  have an isomorphism
\begin{equation}
C^*(a,1_H)^{\ot\infty}
\cong C(\sg(a))^{\ot \infty},
\end{equation}
and hence, on the given identifications, we obtain an isomorphism of C*-algebras
\begin{equation}
C^*(a_1, a_2, \ldots, 1_{H^{\ot\infty}})\cong C(\sg(a)^{\N}). \label{a1a2iso}
\end{equation}
It follows from the definition of the infinite tensor products used here that each state $\om_1$ on $B$ defines a state
$\om_1^{\infty}$ on $B^{\ot\infty}$. Take $B=B(H)$ and restrict $\om_1^{\infty}$, which \emph{a priori} is a state on $B(H^{\ot\infty})$, to its commutative C*-subalgebra $C^*(a_1, a_2, \ldots, 1_{H^{\ot\infty}})$. The isomorphism \er{a1a2iso} then gives a probability measure $\mu_{\ul{a}}$ on the compact space $\sg(a)^{\N}$, where the label $\ul{a}$ now refers to the infinite set of commuting operators $(a_1, a_2, \ldots)$ on $H^{\ot\infty}$. To compute this measure, I use  \er{BornfromGelfand} and the fact that by construction functions of the type
\beq
f(\lm_1, \lm_2, \ldots)=f^{(N)}(\lm_1, \ldots, \lm_N),
\eeq  
 where $N<\infty$ and $f^{(N)}\in C(\sg(a)^N)$, are dense in $C(\sg(a)^{\N})$ (with respect to the appropriate supremum-norm), and that in turn finite linear combinations of factorized functions $f^{(N)}(\lm_1, \ldots, \lm_N)=f_1(\lm_1)\cdots f_N(\lm_N)$ are dense in $C(\sg(a)^{N})$. 
 It follows from this that 
 \beq
 \mu_{\ul{a}}=\mu_a^{\infty}. \label{mula}
 \eeq
 Since this generalizes \er{mulaN} to $N=\infty$, 
  the proof of Theorem \ref{ET} is finished. \hfill $\Box$
 \section{1-Randomness}\label{AR}
  In what follows, the notion of 1-randomness, originally defined by Martin-L\"{o}f in the setting of constructive measure theory, will be explained through an equivalent definition in terms of Kolmogorov complexity.\footnote{For details see 
   Volchan (2002), Terwijn (2016), 
   Diaconis \& Skyrms (2018, Chapter 8), and Eagle  (2019) for starters,  technical surveys  by 
  Zvonkin \& Levin (1970), Muchnik \emph{et al.}\ (1998), Downey \emph{et al.}\ (2006), 
  Gr\"{u}nwald \& Vit\'{a}nyi (2008),  and Dasgupta (2011), and  books by Calude (2002), 
   Li \& Vit\'{a}nyi (2008), Nies (2009),  and  Downey \& Hirschfeldt (2010).
  For history see van Lambalgen (1987, 1996) and   Li \& Vit\'{a}nyi (2008).
For physical applications see e.g.\   Earman (1986),  Svozil (1993, 2018), Calude (2004),   Wolf (2015),
 Bendersky \emph{et al.}\  (2016, 2017), 
  Senno (2017),   Baumeler  \emph{et al.}\ (2017), and
Tadaki (2018, 2019).} We assume basic familiarity with the notion of a computable function $f:\N\raw\N$, which may  be defined  through recursion theory or  Turing machines. 
 
 A \emph{string} is a \emph{finite} succession of bits (i.e.\ zeros and ones).  The length of a string $\sg$ is denoted by $|\sg|$. 
The set of all strings of length $N$ is denoted by $\ul{2}^N$, where  $\ul{2}=\{0,1\}$,  and  
\beq
\ul{2}^*=\bigcup_{N\in\N}\ul{2}^N
\eeq denotes the set of all  strings. The  \emph{Kolmogorov complexity} $K(\sg)$ of $\sg\in\ul{2}^*$ is defined, roughly speaking, as the length of the shortest computer program that prints $\sg$ and then halts. We then say, again roughly, that $\sg$ is 
 \emph{Kolmogorov random} if this shortest program contains all of $\sg$ in its code, i.e.\ if the  shortest computable description of  $\sg$ is $\sg$ itself. 
 
 To make this precise,\footnote{\label{Ear}A Turing machine $T$ is \emph{prefix-free} if its domain $D(T)$ consists of a prefix-free subset of $\ul{2}^*$, i.e., if $\sg\in D(T)$ then $\sg\ta\notin D(T)$ for any $\sg,\ta\in \ul{2}^*$, where $\sg\ta$ is  the concatenation of $\sg$ and $\ta$: if $T$ halts on input $\sg$ then it does not halt on either any initial part or any extension of $\sg$. 
 The prefix-free version is only needed to correctly define randomness of sequences  in terms of randomness of their initial parts,
 which is necessary to satisfy \emph{Earman's Principle}:
`\emph{While idealizations are useful and, perhaps, even essential to progress in physics, a sound principle of interpretation would seem to be that no effect  can be counted as  a genuine physical effect if it disappears
when the idealizations are removed.'}  See Earman (2004), p.\ 191.  For finite strings $\sg$ one may work with the \emph{plain Kolmogorov complexity} $C(\sg)$, defined as the length (in bits) of the shortest computer program (run on some fixed universal Turing machine $U$) that computes $\sg$. }
 fix some  universal prefix-free Turing machine $U$, seen as performing a computation on input $\ta$ (in its prefix-free domain) with output $U(\ta)$,  and
 define   \begin{equation}
K(\sg)=\min_{\ta\in\ul{2}^*} \{|\ta|: U(\ta)=\sg\}.
\end{equation}
The function $K:\ul{2}^*\raw\N$ is uncomputable, but that doesn't mean it is ill-defined.
 The choice of $U$ affects $K(\sg)$  up to a $\sg$-independent constant, and to take this dependency into account we 
 state certain results in terms of the ``big-O'' notation familiar from Analysis.\footnote{Recall that $f(n)=O(g(n))$ iff there are constants $C$ and $N$ such that $|f(n)|\leq C|g(n)|$ for all $n\geq N$.} For example, if $\sg$ is easily computable, like the first $|\sg|$ binary digits of $\pi$, then 
 \beq
 K(\sg)=O(\log|\sg|),
 \eeq
  with the logarithm in base 2 (as only the length of $\sg$ counts). However, a random $\sg$ has 
  \beq
  K(\sg)=|\sg|+ O(\log|\sg|).
  \eeq
 We
 say that $\sg$ is \emph{$c$-Kolmogorov random}, for some $\sg$-independent constant $c\in\N$, if 
 \beq
 K(\sg)\geq |\sg|-c.
 \eeq
  Simple counting arguments show that as $|\sg|=N$ gets large,  the overwhelming majority of strings in $\ul{2}^N$ (and hence in $\ul{2}^*$) is $c$-random.\footnote{It is easy to show that least $2^N-2^{N-c+1}+1$ strings $\sg$ of length $|\sg|=N$ are $c$-Kolmogorov random.} The following theorem, which might be called \emph{Chaitin's first incompleteness theorem},  therefore shows  that randomness is elusive:\footnote{\label{Chaitintheorem}Here ``sound" means that all theorems proved by $T$ are true; this is a stronger assumption than consistency (in fact only the arithmetic fragment of $T$ needs to be sound). One may think of Peano Arithmetic (PA) or of Zermelo--Fraenkel set theory with the axiom of choice (ZFC).
  As in G\"{o}del's  theorems,
  one also assumes that $T$ is formalized  as an axiomatic-deductive system in which proofs could in principle be carried out mechanically by a computer. The status of the true but unprovable sentences $K(\sg)>C$ in Chaitin's theorem is similar to that of the sentence
 $G$ in G\"{o}del's original proof of his first incompleteness theorem, which roughly speaking is an arithmetization of the statement ``I cannot be proved in $T$": assuming soundness and hence consistency of $T$, one can prove 
    $\mathsf{G}$ and $K(\sg)>C$ in the usual interpretation of the arithmetic  fragment of $T$ in the natural numbers $\N$.
   See Chaitin (1987) for his own presentation and analysis of his  incompleteness theorem. 
    Raatikainen (1998) also gives a detailed presentation of the theorem, including a  critique of  Chaitin's ideology.
    Incidentally, he shows that there even exists a $U$ with respect to which $K(\cdot)$ is defined such that $C=0$ in ZFC. 
      See also Franz\'{e}n (2005) and G\'{a}cs (1989).}
\begin{theorem}\label{Chaitin}
For any sound mathematical theory $T$ containing enough arithmetic there is a constant $C\in\N$ such that $T$ cannot prove any sentence of the form $K(\sg)>C$ (although infinitely many such sentences are true), and as such $T$ can only prove (Kolmogorov) randomness of finitely many strings (although infinitely many strings \emph{are} in fact random).
\end{theorem}
 The proof is quite complicated in its details but it is based on the existence of a computably enumerable (c.e.)  list  $\mathsf{T}=(\ta_1,\ta_2,\ldots)$ of the theorems of $T$, and on the fact that after G\"{o}delian encoding by numbers,
    theorems of any given grammatical form can be computably searched for in this list and will eventually be found.  In particular, there exists a program $P$ (running on 
the universal prefix-free Turing machine $U$ used to define $K(\cdot)$) such that $P(n)$ halts iff there exists a string $\sg$
for which $K(\sg)>n$ is a theorem of $T$. If there is such a theorem the output is $P(n)=\sg$, where $\sg$ appears in the first such theorem of the kind (according to the list  $\mathsf{T}$). By definition of  $K(\cdot)$, this means that
\beq
K(\sg)\leq |P|+|n|.
\eeq
 Now suppose that no $C$ as in the above  statement of the theorem exists. Then there is $n\in\N$ large enough that $n>|P|+|n|$ and there is a string $\sg\in\ul{2}^*$ such that $T$ proves $K(\sg)>n$. Since $T$ is sound  this is actually true,\footnote{ The following contradiction can be made more dramatic by taking $n$ such that $n>>|P|+|n|$.} which  gives a contradiction between 
 \begin{align}
 K(\sg)>n>|P|+|n|; && K(\sg)\leq |P|+|n|.
 \end{align}
Note that this proof shows that a \emph{proof} in $T$ of $K(\sg)>n$ (if true) would also \emph{identify} $\sg$.
\smallskip

As an  idealization of a long  (binary) string, a (binary) \emph{sequence} $x=x_1x_2\cdots$ is an infinite succession of bits, 
i.e.\ $x\in \ul{2}^{\N}$, with finite truncations $x_{|N}=x_1\cdots x_N\in \ul{2}^N$ for each $N\in\N$.  We then call $x$  \emph{Levin--Chaitin random} if
each truncation of $x$ is $c$-Kolmogorov random for some $c$, that is, if
 there exists $c\in\N$ such that $K(x_{|N})\geq N - c$ for each $N\in\N$. 
Equivalently,\footnote{See Calude (2002), Theorem 6.38 (attributed to Chaitin) for this equivalence.}
 a sequence $x$ is Levin--Chaitin random if eventually $K(x_{|N})>>N$, in that
\begin{equation}
\lim_{N\raw\infty} (K(x_{|N})-N)=\infty.\label{Calude}
\end{equation}
 Apart from having the same intuitive pull as Kolmogorov randomness (of strings), this definition gains from the fact that it is equivalent to two other appealing notions of randomness, namely
  \emph{patternlessness} and  \emph{unpredictability}, both also defined computationally. 
 
 \noindent  In view of these equivalences we simply call a  Levin--Chaitin random sequence  \emph{1-random}.\footnote{Any pattern in a sequence $x$ would make it compressible, but one has to define the notion of a pattern very carefully  in a computational setting. This was accomplished by Martin-L\"{o}f in 1966, who defined a pattern as a  specific kind of probability-zero subset $T$ of $\ul{2}^{\N}$ (called a ``test")
that can be computably approximated by subsets $T_n\subset \ul{2}^{\N}$ of increasingly small probability $2^{-n}$;
if $x\in T$, then  $x$  displays some pattern and it is patternless iff $x\notin T$ for all such tests.   Martin-L\"{o}f's definition yields what usually called 1-randomness, in view of his use of so-called $\Sg^0_1$ sets.
See the  textbooks  Li \& Vit\'{a}nyi (2008), Calude (2002), Nies (2009), and  Downey \& Hirschfeldt (2010)  for the equivalences between Levin--Chaitin randomness (incompressibility),  Martin-L\"{o}f randomness (patternlessless), and a third notion (unpredictability) that evolved from the work of von Mieses and Ville, finalized by Schnorr.
The name \emph{Levin--Chaitin randomness}, taken from Downey \emph{et al.}\ (2006), is justified by its independent origin in Levin (1973) and Chaitin (1975).}

 A sequence $x\in\ul{k}^{\N}$ is \emph{Borel normal} in base $k$ if each string $\sg$  has frequency $k^{-|\sg|}$ in $x$.
Any hope of defining randomness  as Borel normality  in base 10 is  blocked by \emph{Champernowne's number} $0123456789101112131 \cdots$, which is Borel normal but clearly not random in any reasonable sense (this is also true in base 2).  The
 decimal expansion of $\pi$ is also conjectured to be Borel normal in base 10 (with huge numerical support), although $\pi$ clearly is not random either. However, Borel normality seems a desirable property of truly random numbers on any good definition, and so we are fortunate to have:
\begin{proposition}
A 1-random sequence 
 is Borel normal (in base 2, but in fact in any base) and hence (``monkey typewriter theorem'') contains any finite string infinitely often.\footnote{For details and proofs see Calude (2002), Corollary 6.32 in \S 6.3 and almost all of
 \S 6.4. } 
 \end{proposition}
 Another desirable property comes from the following theorem due to Martin-L\"{o}f, in which $P$ is the 50-50 probability on $\{0,1\}$ and $P^{\infty}$ is the induced probability measure on $\ul{2}^{\N}$:
 \begin{theorem}\label{PML}
 With respect to $P^{\infty}$ almost every outcome sequence $x\in\ul{2}^{\N}$ is 1-random.
  \end{theorem}
 This implies that the 1-random sequences form an uncountable subset of $\ul{2}^{\N}$,\footnote{To see this, use the measure-theoretic isomorphism between $(\ul{2}^{\N},\Sigma_K,P^{\infty})$
 and $([0,1],\Sg_L,dx)$, where $\Sg_K$ is the ``Kolmogorov'' $\sg$-algebra generated by the cylinder sets
 $[\sg]=\{x\in \ul{2}^{\N}\mid x_{||\sg|}=\sg\}$, where $\sg\in\ul{2}^*$, and  $\Sg_K$ is the ``Lebesgue'' $\sg$-algebra
generated by the open subsets of $[0,1]$. See also Nies (2009), \S 1.8. } although  topologically this subset is meagre (i.e.\ Baire first category).\footnote{See Calude (2002), Theorem 6.63. Hence meagre subsets of $[0,1]$ exist with unit Lebesgue measure!
}
 Chaitin's (first) incompleteness theorem for (finite) strings has the following counterpart for (infinite) sequences:
\begin{theorem}\label{Klaas}
If $x\in\ul{2}^{\N}$ is 1-random, then ZFC (or any  sufficiently comprehensive mathematical theory $T$ meant in Theorem \ref{Chaitin})
can compute only finite many digits of $x$.\footnote{More precisely, only finitely many true statements of the form: `the $n$'th bit $x_n$ of $x$ equals its actual value' (i.e.\ 0 or 1) are provable in $T$ (where a proof in $T$ may be seen as a computation, since one may algorithmically search for this proof in a list). See Calude (2002), Theorem 8.7, which is stated for Chaitin's $\Omega$ but whose proof holds for any 1-random sequence. Indeed, as pointed out to the author by Bas Terwijn, even more generally, ZFC (etc.) can only compute finitely many digits of any \emph{immune} sequence (we say that a sequence $x\in\ul{2}^{\N}$ is \emph{immune}
if the corresponding subset $S\subset\N$ (i.e.\ $1_S=x$) contains no infinite c.e.\ subset), and by (for example) Corollary 6.42 in Calude (2002) any 1-random sequence is immune. 
 }
\end{theorem}
This clearly excludes defining a 1-random number by somehow listing its digits, but some can be described by a formula. One example is Chaitin's $\Omega$, or more precisely $\Om_U$,\footnote{There  exists a $U$ for which not a single digit of $\Om_U$ can be known, see  Calude (2002), Theorem 8.11.}
 which is the halting probability of some fixed universal prefix-free Turing machine $U$, given by
\begin{equation}
\Om_U:=\sum_{\ta\in \ul{2}^*\mid U(\ta)\downarrow}2^{-|\ta|}.
\end{equation}
\section{Bell's theorem and  free will theorem}\label{FWT}
In support of the analysis of hidden variable theories in the main text, this appendix  reviews Bell's (1964) theorem and the free will theorem, streamlining earlier expositions  (Cator \&  Landsman, 2014;  Landsman, 2017, Chapter 6) and leaving out proofs and other adornments.\footnote{The original reference for Bell's theorem is Bell (1964); see further footnote \ref{Bellfn}, and in the context of this appendix also Esfeld (2015) and Sen \& Valentini (2020) are relevant.
 The free will theorem originates in  Heywood \& Redhead (1983), followed by Stairs (1983),
 Brown \& Svetlichny (1990),  Clifton (1993), and, as name-givers, Conway \& Kochen (2009). Both theorems can and have been presented and interpreted in many different ways, of which we choose the one that is relevant for the general discussion on randomness in the main body of the paper. This appendix is taken almost \emph{verbatim} from Landsman (2020). } 
 In the specific context of 't Hooft's theory (where the measurement settings are determined by the hidden state) and Bohmian mechanics (where they are not, as in the original formulation of Bell's theorem and in most hidden variable theories) an advantage of my approach is that both free (uncorrelated) und correlated settings fall within its scope; the former  are distinguished from the latter by  an independence assumption.\footnote{
  This addresses a problem Bell faced even according to some of his most ardent supporters
(Norsen, 2009; Seevinck \& Uffink, 2011), namely the tension between the idea that the hidden variables (in the pertinent causal past) should on the one hand include all ontological information relevant to the experiment, but on the other hand should leave Alice and Bob free to choose any settings they like. Whatever its ultimate fate, 't Hooft's staunch determinism has drawn attention to issues like this, as has the free will theorem.}

As a warm-up I start with a version of the Kochen--Specker theorem, whose logical form is very similar to Bell's (1964) theorem and the free will theorem, as follows:
    \begin{theorem}\label{KSthm}
 Determinism, {\sc qm}, non-contextuality, and free choice  are contradictory.
\end{theorem}
Of course, this unusual formulation hinges on the precise meaning of these terms. 
\begin{itemize}
\item 
\hi{determinism} is the conjunction of the following two assumptions.

1. There is a state space $X$  with associated functions $A: X\raw S$ and $L:X\raw O$,
where $S$ is  the set of all  possible \emph{measurement settings} Alice can choose from, namely 
 a suitable finite set of orthonormal bases of $\R^3$ (11 well-chosen bases will do to arrive at a contradiction),\footnote{If her setting is a basis 
 $(\vec{e}_1,\vec{e}_2,\vec{e}_3)$, Alice measures the quantities $(J_{\vec{e}_1}^2, J_{\vec{e}_2}^2, J_{\vec{e}_3}^2)$, where
 $J_{\vec{e}_1}=\la\vec{J},\vec{e}_i\ra$ is the component of the angular momentum operator $\vec{J}$ of a massive spin-1 particle in the direction $\vec{e}_i$.} and  $O$ is some set of possible  \emph{measurement outcomes}. Thus some $x\in X$ determines \emph{both} Alice's setting $a=A(x)$ \emph{and} her 
 outcome  $\al=L(x)$. 

2. There exists some set $\Lm$ and an additional function 
$H:X\raw \Lm$ such that  
\beq
L=L(A,H),
\eeq
 in the sense that
 for each $x\in X$ one has $L(x)=\hat{L}(A(x),H(x))$
  for a certain function  $\hat{L}:S \x \Lm\raw O$.  This self-explanatory assumption just states that each measurement outcome 
  $L(x)=\hat{L}(a,\lm)$ 
  is determined by the  measurement setting $a=A(x)$ and the ``hidden" variable or state $\lm=H(x)$  of the particle
 undergoing measurement.
  \item  {\sc qm}  fixes 
 $O=\{(0,1,1), (1,0,1), (1,1,0)\}$, which is a non-probabilistic fact of \qm\ with overwhelming (though  indirect) experimental support. 
   \item \hi{non-contextuality}
stipulates that the function $\hat{L}$ just introduced take the form
  \beq
  \hat{L}((\vec{e}_1,\vec{e}_2,\vec{e}_3),\lm)=(\til{L}(\vec{e}_1,\lm), \til{L}(\vec{e}_2,\lm), \til{L}(\vec{e}_3,\lm)), \label{hatL3}
  \eeq
  for a single function $\til{L}:S^2\x \Lm\raw\{0,1\}$ that also satisfies 
$\til{L}(-\vec{e},\lm)=\til{L}(\vec{e},\lm)$.\footnote{Here $S^2=\{(x,y,z)\in\R^3\mid x^2+y^2+z^2=1\}$ is the 2-sphere, seen as the space of unit vectors in $\R^3$.
Eq.\ \er{hatL3} means that the outcome of Alice's measurement 
of $J_{\vec{e}_i}^2$ is independent of the ``context" $(J_{\vec{e}_1}^2, J_{\vec{e}_2}^2, J_{\vec{e}_3}^2)$; she might as well measure $J_{\vec{e}_i}^2$ by itself. The last equation is trivial, since $(J_{-\vec{e}_i})^2=(J_{\vec{e}_i})^2$.
}
  \item \hi{free choice} finally states that the following function is surjective:
\begin{align}
A\x H:X\raw S\x \Lm; && x\mapsto (A(x),H(x)).
\end{align}
 In other words, 
 for each $(a,\lm)\in S\x\Lm$ there is an $x\in X$  for which $A(x)=a$ and $H(x)=\lm$. This makes   $A$ and $H$  ``independent"\
 (or: makes $a$ and $\lm$ free variables).
  \end{itemize}
See Landsman (2017), \S6.2 for a proof of the Kochen--Specker theorem in this language.\footnote{
The assumptions imply the existence of a coloring $C_{\lm}: \mathcal{P}\raw\{0,1\}$ of $\R^3$, where 
$\CP\subset S^2$ consist of all unit vectors contained in all bases in $S$, and $\lm$ ``goes along for a free ride".
A coloring of $\R^3$ is a function $C:\mathcal{P}\raw \{0,1\}$ such that for any set $\{e_1,e_2,e_3\}$ in  $\mathcal{P}$ with
$e_ie_j=\dl_{ij} 1_3$ and $e_1+e_2+e_3=1_3$ where $1_3$ is the $3\x 3$ unit matrix) there is exactly one  $e_i$ for which $C(e_i)=1$.
 Indeed, one finds
$C_{\lm}(\vec{e})=\til{L}(\vec{e},\lm)$. The key to the proof of Kochen--Specker is that on a suitable choice of the set $S$ such a coloring cannot exist. 
}
\smallskip

\noindent Bell's (1964) theorem and the free will theorem both take a similar generic form, namely:
    \begin{theorem}\label{FWTthm}
Determinism, {\sc qm}, local contextuality, and free choice,   are contradictory.
\end{theorem}
Once again, I have to explain what these terms exactly mean in the given context. 
\begin{itemize}
\item \hi{determinism} is a straightforward adaptation of the above meaning to the bipartite ``Alice and Bob" setting. Thus we have a state space $X$  with associated functions 
\begin{align}
A: X\raw S; & &
B: X\raw S; &&
 L:X\raw O & &
 R: X\raw O, \label{C4}
\end{align}
where $S$, the set  of all possible measurement settings Alice and Bob can each choose from, differs a bit between the two theorems:
for the free will theorem it is the same as for the Kochen--Specker theorem above, as is the set  $O$ of possible measurement outcomes, 
whereas for Bell's theorem (in which Alice and Bob each measure a 2-level system), $S$ is some finite set of angles (three is enough), and 
 $O=\{0,1\}$.
 \begin{itemize}
\item In the free will case, these functions and the state $x\in X$ determine both the settings $a=A(x)$ and $b=B(x)$ of a measurement  and its outcomes  $\al=L(x)$ and $\beta=R(x)$ for Alice on the \emph{L}eft and for Bob on the \emph{R}ight, respectively.
\item  All of  this is also true in the Bell case, but since his theorem relies on impossible measurement statistics (as opposed to impossible individual outcomes),  one in addition assumes a probability measure $\mu$ on $X$.\footnote{The existence of $\mu$ is of course predicated on $X$ being a measure space with corresponding $\sg$-algebra of measurable subsets, with respect to which all functions in \er{C4} and below are measurable.}
\end{itemize}
Furthermore, there exists some set $\Lm$ and some  function  $H:X\raw \Lm$ such that
 \begin{align}
 L=L(A,B,H); &&
 R=R(A,B,H),
\end{align}
 in the sense that for each $x\in X$ one has  functional relationships
 \begin{align}
 L(x)=\hat{L}(A(x),B(x),H(x)); &&
 R(x)=\hat{R}(A(x),B(x),H(x)),\label{GhatG}
\end{align}
  for certain functions
  $\hat{L}:S \x S\x \Lm\raw O$ and  $\hat{R}:S \x S\x \Lm\raw O$. 
  \item \hi{{\sc qm}}  reflects elementary \qm\ of correlated 2-level and 3-level quantum systems for the Bell and the free will cases, respectively, as follows:\footnote{In Bell's theorem quantum theory can be replaced by experimental support (Hensen  \emph{et al.}, 2015). }
   \begin{itemize}
\item  In the \emph{free will theorem},  $O=\{(0,1,1), (1,0,1), (1,1,0)\}$ is the same as for the Kochen--Specker theorem. In addition  \emph{perfect correlation} obtains: if $a=(\vec{e}_1,\vec{e}_2,\vec{e}_3)$ is Alice's orthonormal basis
and  $b=(\vec{f}_1,\vec{f}_2,\vec{f}_3)$ is Bob's, one has
 \begin{equation}
\vec{e}_i=\vec{f}_j\: \Raw\: \hat{L}_i(a,b,z)=\hat{R}_j(a,b,z), \label{ienj}
\end{equation}
where $\hat{L}_i, \hat{R}_j: S \x S\x \Lm\raw \{0,1\}$ are the components of $\hat{L}$ and $\hat{R}$, respectively.  Finally,\footnote{As in Kochen--Specker, this is because Alice \& Bob measure \emph{squares} of (spin-1) angular momenta.} if $(a',b'$) differs from $(a,b)$ by changing the sign of any  basis vector,
 \begin{align}
\hat{L}(a',b',\lm)=\hat{L}(a,b,\lm); && 
\hat{R}(a',b',\lm)=\hat{R}(a,b,\lm).
\end{align}
\item In \emph{Bell's theorem}, $O=\{0,1\}$, and the statistics for the experiment is reproduced as conditional joint probabilities given by the measure $\mu$ through
\begin{equation}
P(L\neq R|A=a,B=b)=\sin^2(a-b).\label{uitkomstAspect2} 
\end{equation}
\end{itemize}
\item \hi{local contextuality}, which replaces and weakens non-contextuality, means that 
\begin{align}
L(A,B,H)=L(A,H); && R(A,B,H)=G(B,H).
\end{align} In words: Alice's outcome \emph{given $\lm$} does not depend on Bob's setting, and \emph{vice versa}. 
\item  \hi{free choice} is an independence assumption that looks differently for both theorems:
\begin{itemize}
\item In the \emph{free will theorem} it means  that  each $(a,b,\lm)\in S\x S\x \Lm$ is possible in that there is an $x\in X$  for which $A(x)=a$, $B(x)=b$, and $H(x)=\lm$.
\item In  \emph{Bell's theorem},  $(A,B,H)$ are \emph{probabilistically independent} relative to $\mu$.\footnote{By definition, this also implies  that the pairs $(A,B)$, $(A,H)$, and $(B,H)$ are also independent.}
\end{itemize}
  \end{itemize}
This concludes the joint statement of the free will theorem and Bell's (1964) theorem in the form we need for the main text. The former is proved by reduction to the Kochen--Specken theorem, whilst the latter follows by reduction to the usual version of Bell's theorem via the free choice assumption; see Landsman (2017), Chapter 6 for details. 

For our purposes these theorems are equivalent, despite subtle differences in their assumptions. Bell's theorem is much more robust in that it does not rely on perfect correlations (which are hard to realize experimentally), and in addition it requires almost no input from quantum theory.
On the other hand,  Bell's theorem uses  probability theory in a highly nontrivial way: like the hidden variable theories it is supposed to exclude it relies on the possibility of fair sampling of the probability measure $\mu$. The 
 factorization condition defining probabilistic independence passes this requirement of fair sampling on to both the hidden variable and the settings, which brings us back to the main text.
 
Different parties may now be identified by the assumption they drop:
 Copenhagen \qm\ rejects determinism,  Valentini (2019) rejects the Born rule and hence {\sc qm},
 Bohmians rejects  local contextuality, and finally
't Hooft rejects free choice. 
However, as we argue in the main text, even the latter two camps do not really have a deterministic theory underneath \qm\ because of their need to randomly sample the probability measure they must use to recover the predictions of \qm. 
\newpage
\addcontentsline{toc}{section}{References}
\begin{small}

\end{small}
\end{document}